\documentclass[journal]{IEEEtran}

\usepackage{graphicx}
\usepackage[dvipsnames]{xcolor}
\usepackage{amssymb}
\usepackage{amsmath}
\usepackage{optidef}
\usepackage{array}
\usepackage{algpseudocode}
\usepackage{multirow}
\usepackage{algorithm}
\usepackage{cite}
\usepackage{csquotes}
\usepackage{float}
\usepackage{todonotes}
\usepackage{soul}
\usepackage[hidelinks]{hyperref}
\usepackage{diagbox}
\usepackage[bottom]{footmisc}
\usepackage{lipsum}

\usepackage{booktabs}
\ifCLASSINFOpdf
\else
\fi
\def\NoNumber#1{{\def\alglinenumber##1{}\State #1}\addtocounter{ALG@line}{-1}}
\begin{document}
%
\title{\huge SemSpaceFL: A Collaborative Hierarchical Federated Learning Framework for Semantic Communication in 6G LEO Satellites}
%
%
%

\author{Loc X. Nguyen, Sheikh Salman Hassan,~\IEEEmembership{Member,~IEEE}, Yu Min Park, Yan Kyaw Tun,~\IEEEmembership{Member,~IEEE,}\\ Zhu Han,~\IEEEmembership{Fellow,~IEEE} and Choong Seon Hong,~\IEEEmembership{Fellow,~IEEE}

\thanks{Loc X. Nguyen, Yu Min Park, Choong Seon Hong are with the Department of Computer Science and Engineering, Kyung Hee University, Yongin-si, Gyeonggi-do 17104, Rep. of Korea, e-mail: \{xuanloc088, yumin0906, cshong\}@khu.ac.kr.}
\thanks{Yan Kyaw Tun is with the Department of Electronic Systems, Aalborg University, A. C. Meyers Vænge 15, 2450 København,  e-mail: ykt@es.aau.dk.}
\thanks{Sheikh Salman Hassan is with the Institute for Imaging, Data and Communications, The University of Edinburgh, Edinburgh, EH9 3BF, United Kingdom, e-mail: shassan@ed.ac.uk.}
\thanks{Zhu Han is with the Department of Electrical and Computer Engineering at the University of Houston, Houston, TX 77004 USA, and also with the Department of Computer Science and Engineering, Kyung Hee University, Seoul, South Korea, 446-701, e-mail:  hanzhu22@gmail.com.}

}


\maketitle

\begin{abstract}
The advent of the sixth-generation (6G) wireless networks, enhanced by artificial intelligence, promises ubiquitous connectivity through Low Earth Orbit (LEO) satellites. These satellites are capable of collecting vast amounts of geographically diverse and real-time data, which can be immensely valuable for training intelligent models. However, limited inter-satellite communication and data privacy constraints hinder data collection on a single server for training. Therefore, we propose SemSpaceFL, a novel hierarchical federated learning (HFL) framework for LEO satellite networks, with integrated semantic communication capabilities. Our framework introduces a two-tier aggregation architecture where satellite models are first aggregated at regional gateways before final consolidation at a cloud server, which explicitly accounts for satellite mobility patterns and energy constraints. The key innovation lies in our novel aggregation approach, which dynamically adjusts the contribution of each satellite based on its trajectory and association with different gateways, which ensures stable model convergence despite the highly dynamic nature of LEO constellations. To further enhance communication efficiency, we incorporate semantic encoding-decoding techniques trained through the proposed HFL framework, which enables intelligent data compression while maintaining signal integrity. Our experimental results demonstrate that the proposed aggregation strategy achieves superior performance and faster convergence compared to existing benchmarks, while effectively managing the challenges of satellite mobility and energy limitations in dynamic LEO networks.
\end{abstract}

\IEEEpeerreviewmaketitle

\section{Introduction}

\IEEEPARstart{T}{he} evolution of wireless communication has paved the way in the development of the 6G networks, which aim to provide ubiquitous connectivity, ultra-low latency, and intelligent data processing to support a wide range of emerging applications. According to the International Telecommunication Union (ITU), ubiquitous connectivity, massive communication, and Integrated AI and Communications are three out of six envisioned 6G usage scenarios, further underscoring the importance of intelligent, interconnected infrastructures\cite{10904090}. To meet these ambitious goals, future networks envision the seamless integration of terrestrial, aerial, and satellite infrastructures to overcome the limitations of conventional ground-based systems. i.e., high deployment costs and limited accessibility in remote and underserved regions~\cite{10445382,nguyen2024semantic,nguyen2025contemporary,10266782}. In this regard, LEO satellite networks have emerged as a critical component, which offers global coverage, reduced latency compared to geostationary satellites, and the ability to support applications such as environmental monitoring, disaster response, and the Internet of Things (IoT) services \cite{10198334,qiao2025deepseek,hassan2025enabling}. For the industrial sector, i.e., SpaceX, OneWeb, and Planet Labs, have accelerated the deployment of LEO constellations, significantly enhancing the capabilities of satellite networks.

However, the rapid expansion of satellite-based services has led to a surge in the volume of data generated, which gives rise to several challenges, including limited bandwidth, high communication costs, and strict privacy requirements. To address these issues, \textit{federated learning} (FL) has been proposed as a privacy-preserving paradigm that allows satellites to collaboratively train machine learning (ML) models without sharing raw data~\cite{mcmahan2017communication,9919976,li2020federated,LEO_FL}. In typical FL settings, each satellite trains a local model using onboard data and transmits model updates to a central aggregator for global model synthesis. Nevertheless, the deployment of FL in LEO satellite networks is hindered by several factors, which are given as follows:
\begin{itemize}
    \item \textit{High Mobility and Intermittent Connectivity:} Frequent satellite handovers and dynamic link availability result in outdated updates and transmission delays.
    \item \textit{Communication Bottlenecks:} Centralized aggregation via a single gateway creates congestion and increases the risk of packet loss.
    \item \textit{Limited Contact Duration:} Short-lived satellite-gateway links constrain the time available for transmitting model updates.
\end{itemize}

To alleviate these limitations, we consider HFL where client devices are grouped under local aggregators that perform intermediate model updates before relaying them to a global server~\cite{9148862}. This hierarchical structure reduces communication load and improves scalability. However, traditional HFL architectures rely on static, terrestrial edge servers, which makes them ill-suited for the dynamic, resource-constrained LEO environment. Furthermore, most FL and HFL implementations target conventional classification tasks, which may not align with the practical needs of satellite communication systems. To bridge the gap between communication efficiency and intelligent model training in dynamic LEO satellite networks, we propose \textit{SemSpaceFL}, a novel HFL framework that jointly trains semantic communication models. By transmitting only the essential information required for accurate message interpretation, semantic communication significantly reduces bandwidth consumption—an essential benefit in bandwidth-constrained and intermittently connected satellite environments. Our proposed SemSpaceFL features a \textit{two-tier aggregation architecture} that enhances scalability and robustness as follows:
\begin{itemize}
    \item \textit{Satellite-to-Gateway Aggregation:} Satellites locally train models and transmit updates to regional terrestrial gateways, which perform intermediate aggregation by considering satellite mobility, energy availability, and link quality.
    \item \textit{Gateway-to-Cloud Aggregation:} Aggregated updates from multiple gateways are relayed to a global cloud server, which synthesizes the final model and coordinates global learning.
\end{itemize}

A distinctive component of SemSpaceFL is its \textit{dynamic aggregation strategy}, which weighs each satellite's contribution based on contextual factors such as mobility, data quality, energy status, and connectivity duration. Furthermore, the framework leverages this hierarchical structure to train \textit{semantic encoding-decoding models}, thereby enabling intelligent data compression while preserving interpretability and accuracy. The main contributions of our work are as follows:
\begin{itemize}
    \item We propose \textbf{SemSpaceFL}, an HFL framework designed to train DJSCC-based semantic communication models using the scattered data from LEO satellites. In this framework, satellites operate as learning agents, while regional gateways perform intermediate aggregation of local models. These sub-region models are then further aggregated at a central cloud server to produce the final global model. This architecture addresses the key challenges of bandwidth limitations, data privacy, and intermittent connectivity in 6G satellite networks.
    \item Secondly, satellite mobility is explicitly accounted for in the framework to ensure reliable communication links. In particular, we formulate an optimization problem to determine the optimal satellite-to-gateway associations across the network, aiming to maximize the local training epochs for each satellite while ensuring that model updates can be sent back to the gateway before the satellite exits the coverage region.
    \item Thirdly, we develop a novel two-tier aggregation mechanism that further enhances model convergence and system scalability by dynamically adjusting satellite contributions to the training process based on the number of training samples, the achievable training epochs, and the quality of updates.
    \item We conduct extensive simulations that demonstrate the superiority of SemSpaceFL over conventional FL and HFL approaches in terms of model accuracy, convergence speed, and communication efficiency.
\end{itemize}

The remainder of this paper is structured as follows: Section~\ref{Related} reviews related work. Section~\ref{System} introduces the system model and outlines the proposed framework. Section~\ref{Proposed} details the semantic communication implementation within the HFL context. Section~\ref{Performance} presents performance evaluations, and finally Section~\ref{Conclusion} concludes the paper.
\vspace{-0.15in}
\section{Related works}\label{Related}

\subsection{FL in Satellite}
\label{FLLEOSat}
In this section, we provide a summary of the literature on FL in satellite networks. The authors in \cite{9674028} were the first to explore the integration of the FL framework into LEO Satellite Constellations, where a model is trained to perform a classification task. They recognized that the communication link between the satellite and the ground server is available only for a limited time window, leading to the proposal of a specialized asynchronous FL algorithm, \textit{FedSat}, to accommodate the network's unique properties.
On the other hand, authors in \cite{so2022fedspace} focused on the convergence rate of the FL algorithm and formulated an optimization problem to speed up the learning process and reduce the training time.
The proposed \textit{FedSpace} algorithm actively schedules global model aggregation based on the deterministic and time-varying connectivity according to satellite orbits. 
An interesting idea had been proposed by authors in \cite{10121575}, where intra-orbit inter-satellite links (ISL) are leveraged to relay the trained model through satellites before reaching the gateway server.
Developing from the idea,  authors in \cite{10409275} designed a structured Satellite FL with the ISL that can speed up the convergence rate and reduce the communication overhead for the in-network aggregation.
Different from the previous works, \cite{10216376} designed a framework to support various on-orbit resource-intensive training tasks by utilizing the collaboration among satellites. Specifically, their FedLEO collects the updated local models through inter-satellite communication and aggregates the global model within a single orbit, eliminating the need for a central server. High-altitude platforms (HAPs) are leveraged to aggregate the models from the satellites instead of a ground server \cite{10039157}. Later on, the authors leveraged different kinds of communication links, i.e., ISL, satellite-HAP link, and inter-HAP links, to accelerate the FL model convergence.

The work in \cite{jiang2024federated} combined the FL with split learning to satellite-terrestrial integrated networks to analyze the sequential data, which was later applied to detect electricity theft. Instead of considering the data at the LEO satellite side, authors of \cite{han2024cooperative} considered it located at mobile devices and presented a collaboration between ground users and the satellites, where it can train models locally or offload training data to the satellites. Additionally, they further proposed the model transmission among satellites within the LEO orbit and the intra-cluster model aggregation. A weight quantization is proposed in \cite{10415259} to improve bandwidth utilization for the transmission of the parameter during the training process. Due to the low contact time between the satellite and the gateway, \cite{10716798} proposed two aggregation algorithms named \textit{sub-structure} and \textit{pseudo-synchronous} for visible and invisible satellites, respectively. Instead of using a ground station for the model aggregation, the model is transmitted to nearby satellites in the same or even different orbits for the aggregation process \cite{10615752}. A mixed combinatorial optimization problem is formulated to minimize the system energy by determining the routing strategy and communication resources. Similarly, model prioritization queues are proposed to join the FL training process due to the scarcity of satellite-ground bandwidth \cite{zhang2024satfed}.
\vspace{-0.1in}
\subsection{HFL for LEO Satellites}
\label{HFLEOSat}
In the work \cite{pei2024cost}, the authors presented a hierarchical federated learning, where the satellites operate as edge devices and aggregate the local model from Internet of Things (IoTs) devices due to their massive coverage. After the edge aggregation process at satellites, they are relayed to the satellite with the direct link ground gateway for the cloud aggregation. Similarly, \cite{10745225} proposed an integrated hierarchical federated learning framework for the space-air-ground integrated networks, which include ground users as training agents and UAVs as the access points. The objective of the proposed problem is to maximize the coverage areas and ensure the fairness of the system. Thus, they proposed deep reinforcement learning to obtain the optimal resource allocation in the networks and also the aggregation weights. An innovative scenario was proposed in \cite{10679111}, where they divided the data into two categories: sensitive and non-sensitive. The non-sensitive data can be freely shared among space-air-ground devices, which is considered offloading the training to other devices. On the other hand, the sensitive is strictly trained at the local site. The collaborations among LEO satellites and GEO satellites are proposed in \cite{10558767}, where the LEO satellites are the training agents, and the GEO satellites are responsible for the model aggregations. Meanwhile, authors in \cite{10699366} also considered hierarchical FL, where one Medium Earth Orbit is considered for the global server aggregation, and the LEO satellites have direct connection links with the MEO satellite aggregate; the model is from the satellite in the same orbit. 
\vspace{-0.1in}
\subsection{Positions in the Literature and Rationale for Proposal}
The studies in Sections \ref{FLLEOSat} and \ref{HFLEOSat} have underscored the advantages of utilizing the available data and the extensive coverage areas of LEO satellites for classification tasks. However, these works do not directly enhance the satellite networks themselves. By training a semantic communication model, we can leverage the bandwidth efficiency and noise robustness properties to improve the data efficiency of the satellite network. Only a few studies address the scenario, such as \cite{gomez2024tackling} explored training an FL framework for the data compression task. Meanwhile, \cite{10445211} trained a semantic communication model using an FL framework with ground users as training clients. In response, we leverage satellite-captured images to train semantic communication, presenting a practical data-driven application. Furthermore, a hierarchical FL framework is proposed to accelerate the training process.

\begin{figure*}[t]
    \centering
    \includegraphics[width=\textwidth, height=3.8in]{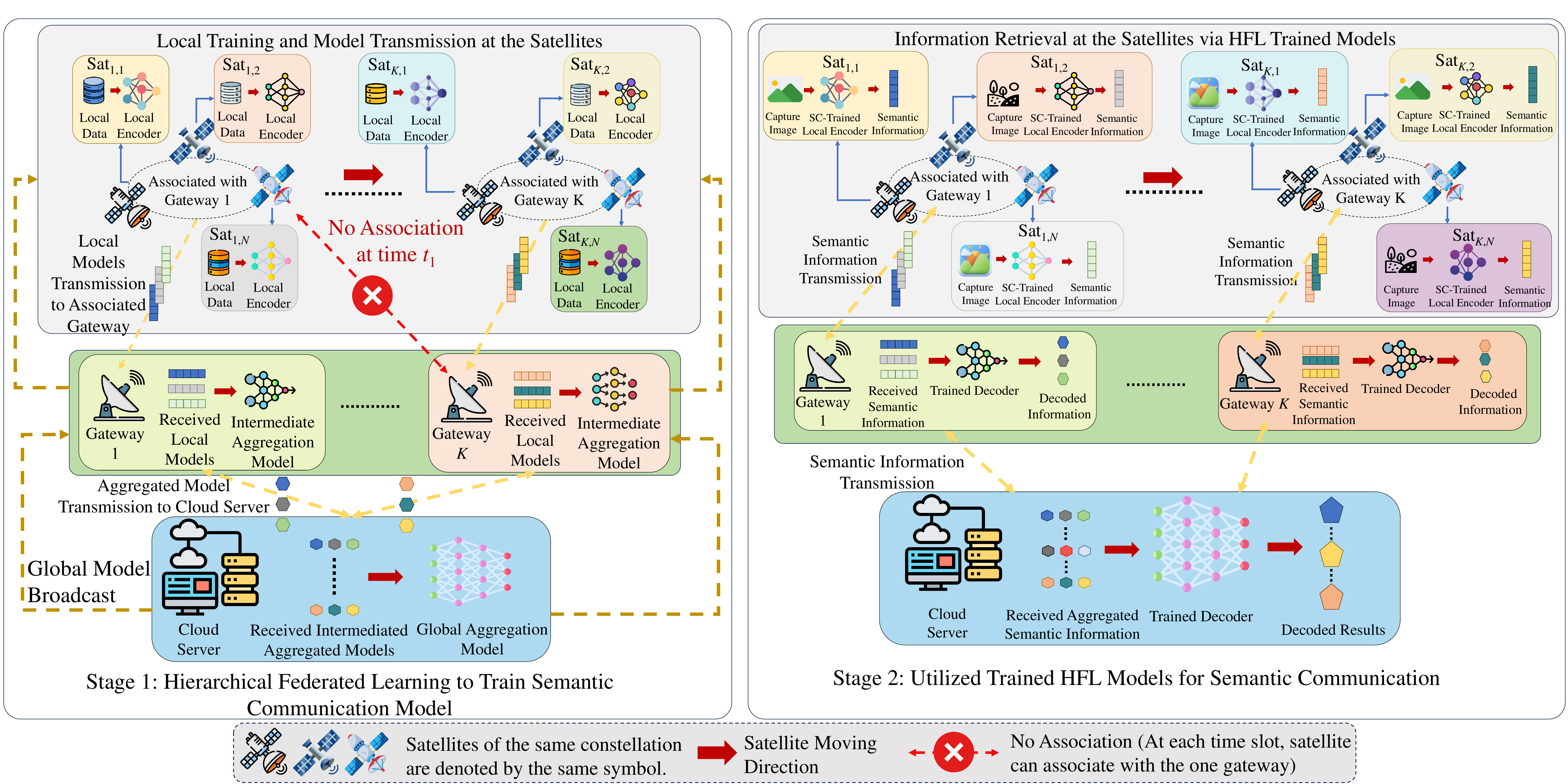}
    \caption{Illustration of collaborative hierarchical federated learning framework for semantic communication in 6G LEO satellites.}
    \label{ImageVisual}
\end{figure*}

\section{System model}\label{System}
We begin by describing the components of the network architecture and its function. As shown in Fig.~\ref{ImageVisual}, the proposed system consists of a set $\mathcal{S}$ of $S$ LEO satellites, a set $\mathcal{G}$ of $G$ gateways, and finally, a single cloud server. Each gateway can cover a portion of the sky and establish communication links with the satellites within its coverage area. However, due to satellite movement, these communication links are highly dynamic and exist only for a limited duration \cite{10901032}. Therefore, we proposed a dynamic HFL framework by leveraging the distributed gateway across geographical areas to train the semantic communication model, which can later be used to provide robustness against the dynamic noise from satellite communication and improve communication efficiency. 
\vspace{-0.1in}
\subsection{HFL training for Cloud-Gateway-Satellites}
In general, the proposed HFL framework simply involves three main mechanisms: local training at the satellite, sub-region aggregation at the gateway, and finally global aggregation at the cloud server. In the sequel, we describe the process from the local training to the higher structure.
\subsubsection{Local training at the satellites} The local model at the satellite $s$ is denoted as $F_{s}^{i,m,k}$, where $i$,$m$,$k$ denote the global, sub-region, and local round, respectively. With the available data here, the local training epoch is presented by:
\begin{equation}
    F_{s}^{i,m,k+1}=  F_{s}^{i,m,k}-\alpha\nabla L_{s}( F_{s}^{i,m,k}),
\end{equation}
where $L_{s}(\cdot)$ is the loss function at the current local training round and $\alpha$ is the learning rate. One noteworthy point is that the number of local training rounds denoted as $K_{s}$, varies at each satellite due to their mobility and is determined by the associated gateway.
\subsubsection{Sub-region Aggregation at the gateway}

After finishing the local training, the satellites send the updated model to the gateways. Here, each gateway aggregates a new sub-region model based on the updated models from the associated satellites, which we refer to as a sub-region round, and the process is mathematically presented by the following equation:
\begin{equation}\label{subregionaggregate}
    F^{i,m+1}_{g}= \sum_{s=1}^{U^{g}}w_{s}^{i,m}F_{s}^{i,m,K}, \forall g \in \mathcal{G},
\end{equation}
where $w_{s}^{i,m}$ denotes the weight contribution of satellite $s$ to the model synthesized during sub-region round $m$. The contribution weight actively changes from each sub-region round, due to the dynamic associated satellites. Here, we use $K$ in (\ref{subregionaggregate}) for simplification, but it is different from satellite to satellite. To start the next sub-region training round, the gateway first determines a new set of associated satellites based on their current position due to satellite mobility. Then, it will determine the number of local training epochs for each satellite based on various factors: position, velocity, distance, and available energy. The detailed information will be provided in Section \ref{gatewaysubsection}.

\subsubsection{Global Aggregation at the Cloud Server}
While the sub-region model is capable of converging to a stationary point, its convergence is relatively slow, and it struggles to achieve a generalized performance due to the limited number of local training satellites. In contrast, the cloud server can efficiently establish connections with distributed gateways, facilitating the aggregation of a global model from each sub-region model. This process synthesizes the learning from diverse and distributed data sources across sub-regions, resulting in a comprehensive and generalized model. The global model aggregation is mathematically presented by:
\begin{equation}
    F^{i+1}=\sum_{g=1}^{G}W^{i}_{g}F^{i,M}_{g}, 
\end{equation}
where $F^{i,M}_{g}$ denotes the model from gateway $g$ at global round $i$, and $W^{i}_{g}$ represents the coefficient of that model. The coefficient reflects the weighted contribution of the gateway model to the aggregation process of the global model at the cloud server. Different from the aggregation at the gateways, the sub-region training round is fixed at $M$ rounds. 
\vspace{-0.1in}
\subsection{Satellite Access Networks}
The massive amount of data collected by a great resource for training deep learning models. However, gathering these enormous amounts of data at one data centre requires many communication resources and violates the data privacy of the satellites. Therefore, in this paper, we examine the satellites that act as federated learning clients. Compared with normal federated learning clients, deploying FL on satellite clients encounters various challenges: 
\begin{itemize}
    \item \emph{Connection Window}: The satellites connect with the gateway for only a short period before moving out of the gateway's coverage range. The disconnection prevents the satellites from transmitting the updated model back to the gateway, which significantly affects the learning process. 
    \item \emph{Heterogeneous Data}: The available data for each satellite varies in amount and distribution properties. In addition, the energy available at the satellites is finite.\\
\end{itemize}
\vspace{-0.2in}
\subsection{Gateways Act as Edge Servers}
\label{gatewaysubsection}
Each gateway will span an area in the sky, and these areas can overlap each other. The gateways will determine the association of satellites in these areas based on their flying directions.
When the gateways receive the global model from the cloud server, it creates connections to all the satellites within its coverage areas and requests information about those satellites: 1) \emph{satellites positions and orbiting direction}, 2) \emph{the number of available data}, 3) \emph{the computing capacity and available energy}. This information will help the gateway determine which satellites should be included in the FL training process and the number of local epochs at the satellite. One noteworthy point is that the satellites receive the training model from which the gateway has to return the training result to that gateway. This established requirement is used to guarantee the stabilization of the training procedure because each training model from gateways is different from the others most of the time, except when they receive the model from the cloud server.
The associated variable: 
\begin{equation}
    \chi^{g}_{s} =\left \{ \begin{array}{ll}{1,} & {\textrm{if the satellite $s$ is associated with the gateway $g$}}  \\ 
      \\{0,} & {\textrm{otherwise.}}\end{array}\right.
\end{equation}
These associated variables significantly impact the training process of each gateway. Moreover, increasing the number of local training epochs typically enhances the overall performance of the sub-region model. The training time of satellite $s$ for one epoch: 
\begin{equation}
    t_{s}= \frac{D_{s}C_{d}}{C_{s}},
\end{equation}
where $D_{s}$ denotes the number of available training data at satellites $s$, while $C_{s}$ is computing frequency that satellite $s$ use for training. $C_{d}$ denotes the number of CPU cycles to train one data sample \cite{10004947}, and it depends on the training model, such as \emph{RestNet, CNN,} and \emph{Transformer}. The energy consumption of satellites used for training can be calculated as follows:
\begin{equation}
        E_{s}={\epsilon}_{s} (C_{s})^{2}C_{d}D_{s},\forall u \in \mathcal{U}_{v},
        \label{energycons}
\end{equation}
where $\epsilon_{s}$ = 5 x $10^{-24}$ is a constant that only depends on the chip architecture mounted on the satellite. In addition, $C_{s}$ denotes the dedicated computing frequency at the satellite to participate in the learning process. The higher computing frequencies can increase the number of local training epochs, consequence in greater energy consumption. However, the energy available on the satellite is limited. Therefore, based on the location, direction, and velocity of each satellite, we can determine how long the satellite will stay in the coverage area of the associated gateway (window time):
\begin{equation}
    T_{s}= \frac{\mathrm{DIS}}{v_{s}},
\end{equation}
where $\mathrm{DIS}$ denotes the remaining distance the satellite will travel while still within the coverage area of gateway $g$, $v_{s}$ denotes the velocity of satellite $s$. With all the association information and the window time for each satellite, gateway $g$ broadcasts its model to all the associated satellites. The satellites utilize their data to train the received models and must transmit the trained models back to the gateways before they move out of coverage areas.

\vspace{-0.1in}
\subsection{Communication Channel Model}

The system considers a heterogeneous antenna configuration. Each satellite is equipped with a single-antenna very small aperture terminal (VSAT), while terrestrial gateways are equipped with large uniform planar arrays (UPAs) that are electronically steerable toward satellite constellations. The UPA at gateway $g$ consists of $N_g = N^x_g \times N^y_g$ antenna elements, where $N^x_g$ and $N^y_g$ denote the number of elements along the horizontal and vertical axes, respectively.

To facilitate scalable and efficient access, gateways employ cooperative orthogonal frequency division multiple access (OFDMA) to simultaneously serve multiple satellites. Specifically, each gateway is capable of forming up to $N^{\textrm{beam}}_g$ independent spot beams, which enables concurrent communication with multiple satellites on orthogonal subcarriers. However, when coverage regions overlap, particularly in edge regions of adjacent beams or across cooperating gateways, \textit{co-channel interference} may arise due to frequency reuse, which needs to be accounted for in the system design.

Assuming a suburban environment with negligible attenuation from rain or clouds, we model the multiple-input single-output (MISO) channel between satellite $s$ and gateway $g$ at time $t$ and subcarrier frequency $f$ as:
\begin{equation}
    h_{s,g} [t,f] = g_{s,g} \cdot \exp\left\{j 2\pi \left(t \nu_{s,g} - f \tau_{s,g} \right)\right\}, \label{channel_model}
\end{equation}
where $g_{s,g}$ represents the complex channel gain incorporating path loss and antenna gains, $\nu_{s,g}$ is the Doppler shift caused by the satellite's high mobility, and $\tau_{s,g}$ is the signal propagation delay between satellite $s$ and gateway $g$. The channel gain $g_{s,g}$ is further expressed as:
\begin{equation}
    g_{s,g} = \sqrt{G_s G_g} \cdot 10^{-\frac{1}{10} \textrm{PL[dB]}},
\end{equation}
where $G_s$ and $G_g$ are the antenna gains of the satellite and gateway, respectively, and $\textrm{PL[dB]}$ denotes the large-scale path loss in decibels. This path loss includes free-space propagation loss, atmospheric gas attenuation, and scintillation effects, which are consistent with recommendations from 3GPP and ITU-R channel modeling guidelines.

\vspace{-0.1in}
\subsection{Communication Link Analysis}
We define a binary association variable \( \chi^s_{g} \in \{0,1\} \) to indicate whether satellite \( s \in \mathcal{S} \) is associated with gateway \( g \in \mathcal{G} \). Specifically, \( \chi^s_{g} = 1 \) implies that satellite \( s \) transmits its local model to gateway \( g \), while \( \chi^s_{g} = 0 \) otherwise. For each gateway \( g \), we define the set of associated satellites as:
\begin{equation}
    \mathcal{S}_g = \{ s \in \mathcal{S} \, | \, \chi^s_{g} = 1 \}.
\end{equation}
The set of interfering satellites not associated with gateway \( g \), but transmitting to other gateways \( g' \in \mathcal{G}, g' \neq g \), is given by:
\begin{equation}
       \mathcal{S}_{\text{intf},g} = \{ s' \in \mathcal{S} \, | \, \exists g' \in \mathcal{G},\, g' \neq g \text{ and } \chi^{s'}_{g'} = 1 \}.
\end{equation}
Each satellite \( s \in \mathcal{S}_g \) applies a beamforming vector \( \boldsymbol{w}_{s,g} \) for transmission to gateway \( g \). The received signal at gateway \( g \) is expressed as:
\begin{equation}
    y_g = \sum_{s \in \mathcal{S}_g} \boldsymbol{h}_{s,g}^{\mathrm{H}} \boldsymbol{w}_{s,g} x_s 
    + \sum_{s' \in \mathcal{S}_{\text{intf},g}} \boldsymbol{h}_{s',g}^{\mathrm{H}} \boldsymbol{w}_{s',g'} x_{s'} 
    + n_g, \label{eq:received_signal_interference}
\end{equation}
where \( \boldsymbol{h}_{s,g} \) is the channel vector from satellite \( s \) to gateway \( g \), \( \boldsymbol{w}_{s,g} \) is the beamforming vector used by satellite \( s \) for gateway \( g \), \( x_s \) is the local model transmitted by satellite \( s \), and \( n_g \sim \mathcal{CN}(0, \sigma^2) \) is the AWGN at gateway \( g \). The signal-to-interference plus noise (SINR) at gateway \( g \) for decoding the signal from satellite \( s \in \mathcal{S}_g \) is given by:
\begin{equation}
    \gamma_{s,g} = \frac{|\boldsymbol{h}_{s,g}^{\mathrm{H}} \boldsymbol{w}_{s,g}|^2}{
    \sigma^2 + \sum\limits_{s' \in \mathcal{S}_{\text{intf},g}} |\boldsymbol{h}_{s',g}^{\mathrm{H}} \boldsymbol{w}_{s',g'}|^2}, \label{eq:sinr_final}
\end{equation}
where each \( s' \in \mathcal{S}_{\text{intf},g} \) is associated with some \( g' \neq g \).

We assume each satellite has a total transmission bandwidth budget \( B^{\mathrm{tot}} \), the bandwidth allocation must satisfy:
\begin{equation}
    \sum_{g \in \mathcal{G}} \chi^s_{g} B_{s,g} \leq B^{\mathrm{tot}}, \quad \forall s \in \mathcal{S},
\end{equation}
where \( B_{s,g} \) is the allocated bandwidth for the \( s \rightarrow g \) link. The achievable transmission rate between satellite \( s \) and gateway \( g \) is:
\begin{equation}
    R_{s,g} = \chi^s_{g} B_{s,g} \log_2 (1 + \gamma_{s,g}).
\end{equation}

The communication time from satellite \( s \) to gateway \( g \) for uploading the local model is:
\begin{equation}
    t^{s \rightarrow g} = \frac{D_{s,g}}{R_{s,g}} + \frac{d_{s,g}}{c},
\end{equation}
where \( D_{s,g} \) is the local model size (in bits), \( d_{s,g} \) is the distance between satellite \( s \) and gateway \( g \), and \( c \) is the speed of light. The downlink communication time for broadcasting the updated global model from gateway \( g \) to satellite \( s \) is:
\begin{equation}
    t^{g \rightarrow s} = \frac{D_{g,s}}{R_{s,g}} + \frac{d_{g,s}}{c}.
\end{equation}

\subsection{Problem Formulation}
The proposed system aims to achieve the best deep learning model with the fewest training rounds on the cloud server by optimizing the following control variables: associate decision, computing frequency, training epochs at the satellites, and finally, optimal aggregation policy at gateways and the cloud server. Therefore, the problem is mathematically described as follows:
\begin{mini!}|s|
    {\boldsymbol{\chi}, \boldsymbol{C}, \boldsymbol{k}, \boldsymbol{w}, \boldsymbol{W}}{
        \frac{1}{S} \sum_{s=1}^{S} \mathbb{E}(L_s)\label{c0}
    }{}{}
    \addConstraint{
        \chi^{s}_{g} \in \{0, 1\}, \quad \forall s \in \mathcal{S}, \quad \forall g \in \mathcal{G}
    } \label{c1}
    \addConstraint{
        \sum_{g=1}^{G} \chi^{g}_{s} \leq 1, \quad \forall s \in \mathcal{S}
    } \label{c2}
    \addConstraint{
        t^{g \rightarrow s} + t_{s} K_{s} + t^{s \rightarrow g} \leq T_{s}, \quad \forall s \in \mathcal{S}
    } \label{c3}
    \addConstraint{
        E_{s} K_{s} \leq E^{m}_{s}
    } \label{c8}
    \addConstraint{
        C_{s} \leq C^{m}_{s}
    } \label{computing}
    \addConstraint{
        \sum_{s=1}^{U^{g}} w_{s} = 1, \quad \forall g \in \mathcal{G}
    } \label{c5}
    \addConstraint{
        0 \leq w_{s} \leq 1, \quad \forall s \in U^{g}, \quad \forall g \in \mathcal{G}
    } \label{c4}
    \addConstraint{
        \sum_{g=1}^{G} W_{g} = 1
    } \label{c7}
    \addConstraint{
        0 \leq W_{g} \leq 1, \quad \forall g \in \mathcal{G},
    } \label{c6}
\end{mini!}
where objective (\ref{c0}) is the average expected loss of the training satellites.  Constraint (\ref{c1}) denotes the constraint of the association variable, whether satellite $s$ is associated with the gateway $g$ or not. Constraint (\ref{c2}) guarantees the satellite is only associated with one gateway at a time, while Constraint (\ref{c3}) limits the local training epochs of each satellite to send back the trained model on time. The value of the weighted contribution of each satellite is limited by (\ref{c4}), and the summation of all the weights in the model aggregation process of gateway $s$ is equal to one as (\ref{c5}). Similarly, Constraints (\ref{c6}) and (\ref{c7}) ensure that the contributions in the federated learning process at the cloud server are properly normalized, requiring their sum to equal one. Finally, Constraint (\ref{c8}) limits the energy consumption of the satellites.
\vspace{-0.1in}
\section{proposed solution}\label{Proposed}

The proposed problem is complicated to solve by a single particular algorithm due to the complexity of different control variables. Therefore, to reduce the computing complexity, we divide the problem into three sub-problems: satellite association and training epochs, federated learning at gateways, and finally, federated learning at the cloud server. Thereafter, we propose individual algorithms to solve each sub-problem and sequentially solve each problem.
\vspace{-0.1in}
\subsection{Satellites Association \& Local Training at Satellites}

First, it is important to determine the satellites associated with one gateway network so that the gateway can broadcast its learning model to those connected satellites. In addition, the number of epochs will also be decided at this stage based on the current position, moving directions, and energy available to the satellites. With the fixed aggregation policies for the federated learning process at the gateways and the cloud server, the first sub-problem is expressed as follows:
\vspace{-0.075in}
\begin{mini!}|l|
	{\boldsymbol{\chi},\boldsymbol{C},\boldsymbol{k}}{\frac{1}{S}\sum_{s=1}^{S}\mathbb{E}(L_{s})}
	{}{}
	\addConstraint{\chi^{s}_{g} \in \{0;1\},  \forall s \in \mathcal{S}, \forall g \in \mathcal{G}}\label{sc1_1}
	\addConstraint{\sum_{g=1}^{G} \chi^{g}_{s}\leq 1, \forall s \in \mathcal{S}}\label{sc1_2}
 	\addConstraint{ t^{g\rightarrow s} + t_{s}K_{s} + t^{s\rightarrow g} \leq T_{s}, \forall s \in \mathcal{S}}\label{sc1_3}
        \addConstraint{E_{s}K_{s} \leq E^{m}_{s}}\label{sc1_4}
                \addConstraint{ C_{s} \leq C^{m}_{s}.}\label{computing_17}
\end{mini!}

However, determining the correlation between the associated variables and the averaged loss value, as well as the computing frequency, is infeasible. On the other hand, increasing the number of local training epochs at each satellite can enhance performance \cite{10158912,10531097,mendieta2022local}. Therefore, rather than directly minimizing the average training loss across satellites, we focus on maximizing the local training epochs for each satellite. Consequently, the problem is reformulated as follows:
\vspace{-0.10in}
\begin{maxi!}|l|
	{\boldsymbol{\chi},\boldsymbol{C}}{\frac{1}{S}\sum_{s=1}^{S}(K_{s})}
	{}{}
	\addConstraint{\chi^{s}_{g} \in \{0;1\},  \forall s \in \mathcal{S}, \forall g \in \mathcal{G}}\label{sc1_1_max}
	\addConstraint{\sum_{g=1}^{G} \chi^{g}_{s}\leq 1, \forall s \in \mathcal{S}}\label{sc1_2_max}
 	\addConstraint{ t^{g\rightarrow s} + t_{s}K_{s} + t^{s\rightarrow g} \leq T_{s}, \forall s \in \mathcal{S}}\label{sc1_3_max}
        \addConstraint{E_{s}K_{s} \leq E^{m}_{s}}\label{sc1_4_max}
                \addConstraint{ C_{s} \leq C^{m}_{s}.}\label{computing_max}
\end{maxi!}
All constraints remain unchanged, while the objective is now changed to maximize the average number of training epochs. We then divide the problem into two sub-problems as follows.

\subsubsection{Satellite Association} 
\begin{figure}[t]
    \centering
    \includegraphics[width=0.49\textwidth, height=2.15in]{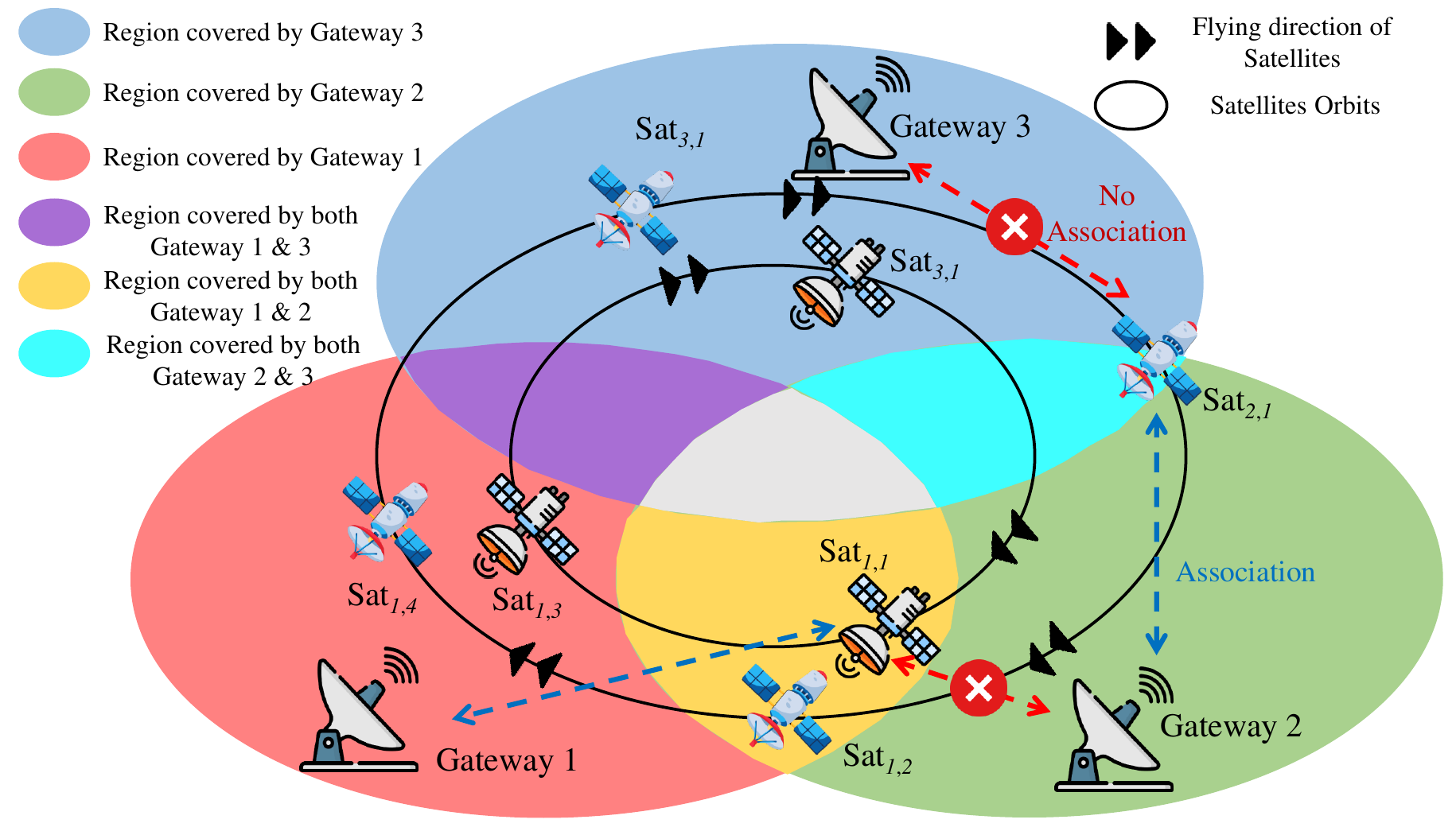}
    \caption{Illustration of the proposed association approach.}
    \label{AssociationSolution}
\end{figure}

It is worth noticing that satellite journeys are pre-defined and fixed ahead of time to avoid collision among satellites within and across constellations. We combine this information along with the current position information of satellites to determine the associated satellites with each gateway. Specifically, we calculate the moving time of the satellite within each sky region covered by the gateway based on the satellite's position and its orbit, and associate it with the gateway region having the maximum transit duration. In Fig.~\ref{AssociationSolution}, we demonstrate the proposal approach for the association problem. Those satellites staying in regions covered by a single gateway will be associated with that particular gateway. The problem is when the satellite is within the collapse region of two gateways; for example, the satellite $Sat_{1,1}$ is closer to the gateway $2$ than $1$; however, due to the flying orbit, the satellite has longer window time to communicate to gateway $1$ rather than gateway $2$. Consequently, associating the satellite $Sat_{1,1}$ to gateway $1$ offers more training epochs and accelerates the convergence of the framework. 
\subsubsection{Computing Frequency} Given the association variable, the problem now becomes maximizing the number of training epochs of each satellite for the local computing frequency while considering the communication and energy constraints. The solution to this problem can be obtained in a closed form as follows:
\vspace{-0.1in}
\begin{align}
    &    C_{s}=\textrm{min} \big{(} C^{m}_{s}; \sqrt[3]{\frac{E^{s}_{m}}{\epsilon_{s} (T_{s}- t^{g\rightarrow s}- t^{s\rightarrow g})}} \space\big{)}, \forall s \in \mathcal{S} \label{ComputingFrequency} \\
    &K_{s}= \frac{E^{m}}{\epsilon_{s} C_{s} C_{d} D_{s}},\forall s \in \mathcal{S}.\label{EpochsEquation}
\end{align}

With the association variables, the dedicated computing frequencies, and the number of training epochs, we broadcast the learning model from the cloud server to gateways and finally to all the satellites to start the training process. By maximizing the number of training epochs of each training client, the convergence of the framework is significantly improved, which leads to an improvement in communication efficiency bottlenecks. The sequence detail is given in the Algorithm~\ref{alg0}. The algorithm has low computational complexity $\mathcal{O} (S \cdot (G +2))$ and is suitable for real-time or large-scale satellite scheduling scenarios.

\newfloat{algorithm}{t}{lop}
\begin{algorithm}
\caption{Satellite Association and Computing Frequency Allocation}\label{alg0}

\textbf{Input:}
\begin{itemize}
    \item $\mathcal{S}$: Set of satellites; $\mathcal{G}$: Set of gateways.
    \item $D_s$: Number of training samples for each satellite $s \in \mathcal{S}$.
    \item $E^m$: Available energy for each satellite.
    \item $\boldsymbol{p}_s$: Coordinates \& orbital parameters of each satellite $s$.
\end{itemize}

\begin{algorithmic}[1]
    \State \textbf{Initial Association:} For the satellite staying in a region covered by a single gateway, associate it with that gateway.
    
    \State \textbf{Travel Time-Based Association:} For satellites that can create connections with multiple gateways, compute their expected traveling time relative to each gateway based on orbital dynamics. Assign the satellite to the gateway offering the longest communication duration.
    
    \State \textbf{Computing Frequency Allocation:} Using the association variables, compute the optimal CPU frequency for each satellite that maximizes the number of training epochs under its energy constraint $E^m$, following Equations~(\ref{ComputingFrequency}) and~(\ref{EpochsEquation}).

\end{algorithmic}

\textbf{Output:}
\begin{itemize}
    \item Satellite-gateway association matrix.
    \item Optimal CPU frequencies per satellite.
    \item Number of local training epochs per satellite.
\end{itemize}
\end{algorithm}

\subsection{Sub-region Global Model Aggregation at Gateways}
The aggregation policy of federated learning at the gateway significantly impacts the convergence rate. Therefore, we need to acquire a good policy, which the gateways have to evaluate the learning model from different perspectives. The second sub-problem is mathematically described as below:
\begin{mini!}|l|
	{\boldsymbol{\boldsymbol{w}}}{\frac{1}{S}\sum_{s=1}^{S}\mathbb{E}(L_{s})}
	{}{}
   	\addConstraint{ 0 \leq w_{s} \leq 1, \forall s \in U^{g}, \forall g \in \mathcal{G} }\label{sc2_1}
       \addConstraint{ \sum_{s=1}^{U^{g}}w_{s} = 1, \forall g \in \mathcal{G} }\label{sc2_2}.
\end{mini!}

With the models being trained at the satellites and sent back to the gateways, the gateway needs to optimize the aggregation process to obtain the highest performance. Unlike previous works that consider the same number of training epochs for each client, we accelerate the training process by adopting different training epochs. Therefore, specifically, our aggregation policy considers three elements: the number of training samples, training epochs, and training loss. These values have high correlations with each other, and they need to be handled carefully to obtain optimal sub-region models. For instance, a satellite with a limited number of data samples might perform multiple local training epochs. However, this does not necessarily imply that the satellite's model provides a greater contribution to the federated learning framework compared to other clients. Taking this argument, we design a new aggregation mechanism for the gateway, where it takes the number of training epochs $K_{s}$ and the number of training samples $D_{s}$ and loss $L_{s}$ as contribution factors into the aggregation process of the gateway region. 
\vspace{-0.05in}
\begin{equation}
    w_{s}=\beta \frac{D_{s}K_{s}^{\kappa}}{\sum_{s=1}^{U^g}D_{s}K_{s}^{\kappa}} + \frac{(1-\beta)}{U_{g}-1}\frac{\sum_{s=1}^{U^g}L_{s}-L_{s}}{\sum_{s=1}^{U^g}L_{s}}, \forall s \in \mathcal{U}^g.\label{subregionaggregation}
\end{equation}
where $\mathcal{U}^{g}$ indicates all the satellites that are associated with the gateway $g$. It should be emphasized that the aggregation is gateway-oriented and not constellation-oriented.
\vspace{-0.05in}
\subsection{Final Global Aggregation at Cloud Server}
At the cloud server, we collect all the trained models from sub-regions and aggregate the final global model from them. 
\begin{mini!}|l|
	{\boldsymbol{W}}{\frac{1}{S}\sum_{s=1}^{S}\mathbb{E}(L_{s})}
	{}{}
        \addConstraint{ 0 \leq W_{g} \leq 1, \forall g \in \mathcal{G}}\label{sc3_1}
        \addConstraint{ \sum_{g=1}^{G}W_{g} = 1.}\label{sc3_2}
\end{mini!}
Similarly to the aggregation process at the gateway, the cloud server determines $W_{g}$ the contribution of the gateway model by evaluating its credibility. The credibility of one gateway is determined by the following equation:
\begin{equation}
    W_{g}=\frac{\sum_{s=1}^{U^g}D_{s}K_{s}^{\kappa}}{\sum_{g=1}^{G}\sum_{s=1}^{U^g}D_{s}K_{s}^{\kappa}}, \forall s \in \mathcal{U^G}, \forall g \in \mathcal{G}.\label{finalglobalmodelaggregation}
\end{equation}
Due to the specialty of the proposed framework, the aggregation process for the global model in the cloud is modified for further refinement and achieves higher performance compared to the conventional case. Algorithm~\ref{alg:Alg1} provides a completed and complete outline of the proposed HFL framework. It improves communication efficiency by reducing direct transmissions to the cloud and enabling local aggregation at gateways. Additionally, this design enhances scalability, allowing the system to support a larger number of satellites without overwhelming the cloud server.

\begin{algorithm}[t]
   \caption{\strut Hierarchical Federated Learning to train a semantic communication model} 
   \label{alg:Alg1}
   \begin{algorithmic}[1]
       \State{\textbf{Initialize:} Global model $\boldsymbol{\Theta}$, number of global epochs $T$, number of gateways $G$.}
        \For{one global round t$=1,2,...,T$}
        \State Broadcast the global model to gateways.
        \State{Solving the Algorithm~\ref{alg0} to get the associated satellites}
        \NoNumber{to each gateway $s\in U^{g}$, satellite training epochs $K_{s}$.}
        \For{each gateway g$=1,2,...,G$}
            \State{Transmit the learning model to the satellite}
          \For{each associated satellite $s\in U^{g}$ $\textbf{in parallel}$}
              \While{Training epoch $k_{s}$ $<$ $K_{s}$}
              \State{Train the model with local data.}
              \State{$\boldsymbol{\Theta}^{k_{s}}_{s} \leftarrow \boldsymbol{\Theta}^{k_{s}-1}_{s}- \eta \nabla \mathcal{L}^{k_{s}}$.}
              \EndWhile
        \EndFor
        \State{Aggregate the gateway model $\boldsymbol{\Theta^g}$ with satellite}
        \NoNumber{contributions are determined by~(\ref{subregionaggregation}).}

            \EndFor
        \State{Aggregate the new global model at the cloud server}
        \NoNumber{from gateway models $\boldsymbol{\Theta^g}$ by following~(\ref{finalglobalmodelaggregation}).}
        \EndFor
    \State{\textbf{Output:} Global Model $\boldsymbol{\Phi}$.}
   \end{algorithmic}
\end{algorithm}

\section{Simulation Settings and Performance Evaluation}\label{Performance}
This section presents a comprehensive set of simulation experiments to evaluate the performance of the proposed \textbf{SemSpaceFL} framework. We assess both system-level efficiency and the acceleration of the federated training process. First, we describe the simulation environment, including communication parameters and learning configurations. Second, we introduce benchmark schemes for comparative analysis. Finally, we provide numerical results that validate the effectiveness of our proposed satellite association and aggregation mechanisms.
\begin{table}[t]
\centering
\setlength{\arrayrulewidth}{0.12mm}
\setlength{\tabcolsep}{8pt}  
\renewcommand{\arraystretch}{1.2}  
\caption{Simulation Parameters}
\label{sim_tab}
\scalebox{0.9}{  
\begin{tabular}{|c|l|c|}
\hline
    \textbf{Notation} & \textbf{Definition} & \textbf{Value} \\ \hline \hline
    $\mathcal{S}$ & Number of Satellites & 10 \\ \hline
    $\mathcal{G}$ & Number of Gateways & 3 \\ \hline
    $C^m_s$ & Max Computing Frequency & 1 GHz \\ \hline
    $C_{d}$ & CPU Requirement for One Sample & $1 \times 10^8$ Cycles \\ \hline
    $G_{g}$ & Gateway Gain & 45 dBi \\ \hline
    $G_{s}$ & Satellite Gain & 25 dBi \\ \hline
    $PL$ & Path Loss & 1.5 dB \\ \hline
    $f$ & Ka-band Carrier Frequency & 10 GHz \\ \hline
    $B_{s,g}$ & Communication Bandwidth & 1 GHz \\ \hline
    $d_{s,g}$ & Satellite Altitude & 500 km \\ \hline    
    $v_{s,g}$ & Doppler Shift & 20 kHz \\ \hline  
    $D_{orb}$ & Radii of Orbits & [1200, 1700, 2200] km \\ \hline  
\end{tabular}}
\end{table}

\subsection{Simulation Setup, Training Configuration, and Dataset Description}
To emulate a realistic LEO satellite network, we consider a topology comprising three ground gateways positioned at coordinates (0; 0; 0), (3,000; 0; 0), and (1,500; 2,000; 0) km. Each gateway is capable of establishing communication links with satellites within a coverage radius of 2,200 km. Furthermore, we simulate three distinct satellite constellations, each centered around the point (1,500, $\frac{2,000}{3}$, 0) km, with orbital radii of 1,200 km, 1,700 km, and 2,200 km, respectively.

To model data heterogeneity across satellites, an essential characteristic in federated learning scenarios, we employ a Dirichlet distribution with concentration parameter \(\lambda = 0.1\) to generate non-identically and independently distributed (non-IID) data splits across satellites. The ImageNet10 dataset is used as the training and evaluation benchmark, with each satellite receiving a disjoint subset.

The goal of training in our framework is to develop a semantic communication model capable of accurately reconstructing original input images from compressed semantic representations. Each satellite trains a local semantic encoder-decoder model using its available dataset. We deploy the Swin Transformer architecture in \cite{yang2023witt} for the deep joint source-channel coding based semantic communication. Energy constraints are introduced by modeling the energy availability of each satellite as a Gaussian random variable with mean \(100\,\text{kJ}\) and standard deviation \(20\,\text{kJ}\). All satellites in the constellations are assumed to move at a constant orbital velocity of \(28,000\,\text{km/h}\). An overview of the simulation parameters is provided in Table~\ref{sim_tab}.

\begin{table*}[t]
\centering
\caption{The performance difference between the proposed association and nearest association under different fl mechanisms.}
\renewcommand{\arraystretch}{1.25} 
\begin{tabular}{|c|cc|cc|cc|cc|}
\hline
Methods & \multicolumn{2}{c|}{\begin{tabular}[c]{@{}c@{}}Nearest Associate \\ FedAvg\end{tabular}} & \multicolumn{2}{c|}{\begin{tabular}[c]{@{}c@{}}Optimal Associate \\ FedAvg\end{tabular}} & \multicolumn{2}{c|}{\begin{tabular}[c]{@{}c@{}}Nearest Associate\\  FedAvep\end{tabular}} & \multicolumn{2}{c|}{\begin{tabular}[c]{@{}c@{}}Optimal Associate \\ FedAvep\end{tabular}} \\ \hline
Metric  & \multicolumn{1}{c|}{PSNR}                             & MS-SSIM                          & \multicolumn{1}{c|}{PSNR}                             & MS-SSIM                          & \multicolumn{1}{c|}{PSNR}                              & MS-SSIM                          & \multicolumn{1}{c|}{PSNR}                              & MS-SSIM                          \\ \hline
SNR=1   & \multicolumn{1}{c|}{25.9789}                          & 0.8610                           & \multicolumn{1}{c|}{26.2396}                          & 0.8691                           & \multicolumn{1}{c|}{26.2895}                           &0.8702                        & \multicolumn{1}{c|}{\textbf{26.4792}}                           & \textbf{0.8754}                        \\ \hline
SNR=3   & \multicolumn{1}{c|}{26.7865}                            & 0.8924                           & \multicolumn{1}{c|}{27.0906}                          & 0.8993                           & \multicolumn{1}{c|}{27.1103}                           & 0.8995                         & \multicolumn{1}{c|}{\textbf{27.4338}}                           & \textbf{0.9056}                       \\ \hline
SNR=5   & \multicolumn{1}{c|}{27.4116}                          & 0.9142                          & \multicolumn{1}{c|}{27.7784}                          & 0.9203                          & \multicolumn{1}{c|}{27.7654}                           &0.9204                          & \multicolumn{1}{c|}{\textbf{28.1848}}                            & \textbf{0.9257}                         \\ \hline
SNR=7   & \multicolumn{1}{c|}{27.8687}                          & 0.9290                           & \multicolumn{1}{c|}{28.2838}                          & 0.9342                     & \multicolumn{1}{c|}{28.2574}                           & 0.9348                        & \multicolumn{1}{c|}{\textbf{28.7330}}                           & \textbf{0.9388}                         \\ \hline
SNR=9   & \multicolumn{1}{c|}{28.1822}                          &0.9384                      & \multicolumn{1}{c|}{28.6300}                          & 0.9429                    & \multicolumn{1}{c|}{28.6037}                           & 0.9441                        & \multicolumn{1}{c|}{\textbf{29.1072}}                           & \textbf{0.9470}                       \\ \hline
SNR=11  & \multicolumn{1}{c|}{28.3842}                          & 0.9442                     & \multicolumn{1}{c|}{28.8552}                          & 0.9480                      & \multicolumn{1}{c|}{28.8322}                           &0.9497                          & \multicolumn{1}{c|}{\textbf{29.3452}}                           & \textbf{0.9520}                         \\ \hline
\end{tabular}
\label{AssociateTable}
\end{table*}

\subsection{Benchmarks for Performance Comparison}
To evaluate the effectiveness of the proposed framework, we compare its performance against several classical benchmark schemes. These benchmarks represent traditional methods in satellite communication and federated learning. The key benchmarks are as follows:
\begin{itemize}
    \item \textbf{Single Gateway FL:} In this benchmark, only a single gateway is deployed. The gateway broadcasts the learning model to satellites for training, and the Federated Averaging (FedAvg) algorithm is employed for model aggregation. This scheme serves as a baseline for comparing the performance of a centralized approach.
    
    \item \textbf{Nearest Association:} This benchmark extends coverage by deploying multiple gateways. Satellites are assigned to the gateway that is geographically closest to them. This scheme aims to improve data utilization while maintaining simplicity in the association process.
    
    \item \textbf{HFL with Various Aggregation Approaches:} In this scenario, multiple gateways are deployed, and satellite association is determined based on satellite position and orbital direction. Different aggregation schemes are considered, including both cloud-based and gateway-based aggregation, to explore their impact on performance in a hierarchical federated learning setup.
\end{itemize}
\subsection{Performance Metrics}
In the paper, we consider the image reconstruction task, whose target is to recover the image as closely as possible to the original under the effects of wireless channel noise. Therefore, we adopted two popular metrics: peak signal-to-noise ratio (PSNR) and multi-scale structural similarity index measure (MS-SSIM) to evaluate the difference between the original and the reconstructed one. The PSNR value is determined by the following equations:
\begin{equation}
    \textrm{PSNR}= 10\log_{10}\frac{\textrm{MAX}^{2}}{\textrm{MSE}},
    \label{PSNRreverseMSE}
\end{equation}
where $\textrm{MAX}$ presents the maximum pixel value of the image \cite{bourtsoulatze2019deep},  the $\textrm{MSE}$ indicates mean squared error between images. In general, PSNR compares images based on pixel values, while MS-SSIM evaluates image quality using multiscale structural similarity; more details are presented in \cite{wang2003multiscale}.

\subsection{Results Analysis}
\begin{figure}[t]
    \centering
    \includegraphics[width=0.5\textwidth]{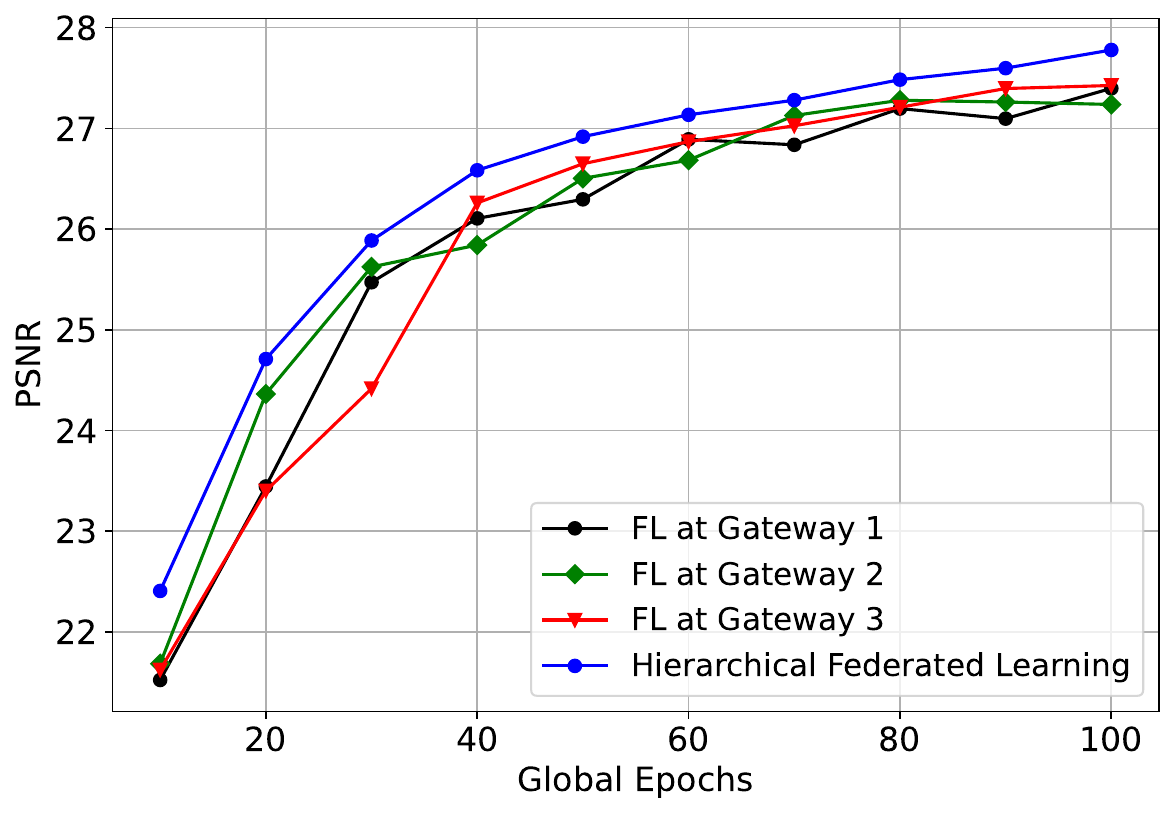}
    \caption{The performance convergence of FL under single Gateway and the collaboration scenarios.}
    \label{HFLPerformance}
\end{figure}

\subsubsection{Performance Enhancement via HFL Adoption}
In this section, we compare the performance of federated learning when each gateway conducts training independently versus when collaboration across gateways is facilitated by a cloud server, as shown in Fig.~\ref{HFLPerformance}. In all cases, we adopt average sample aggregation, and the system operates under a wireless noise level of 5 dB. The satellite association solutions are based on the scheme proposed in this work. 
\setcounter{figure}{4}
\begin{figure*}[t!]
    \centering
    \includegraphics[width=0.9\textwidth]{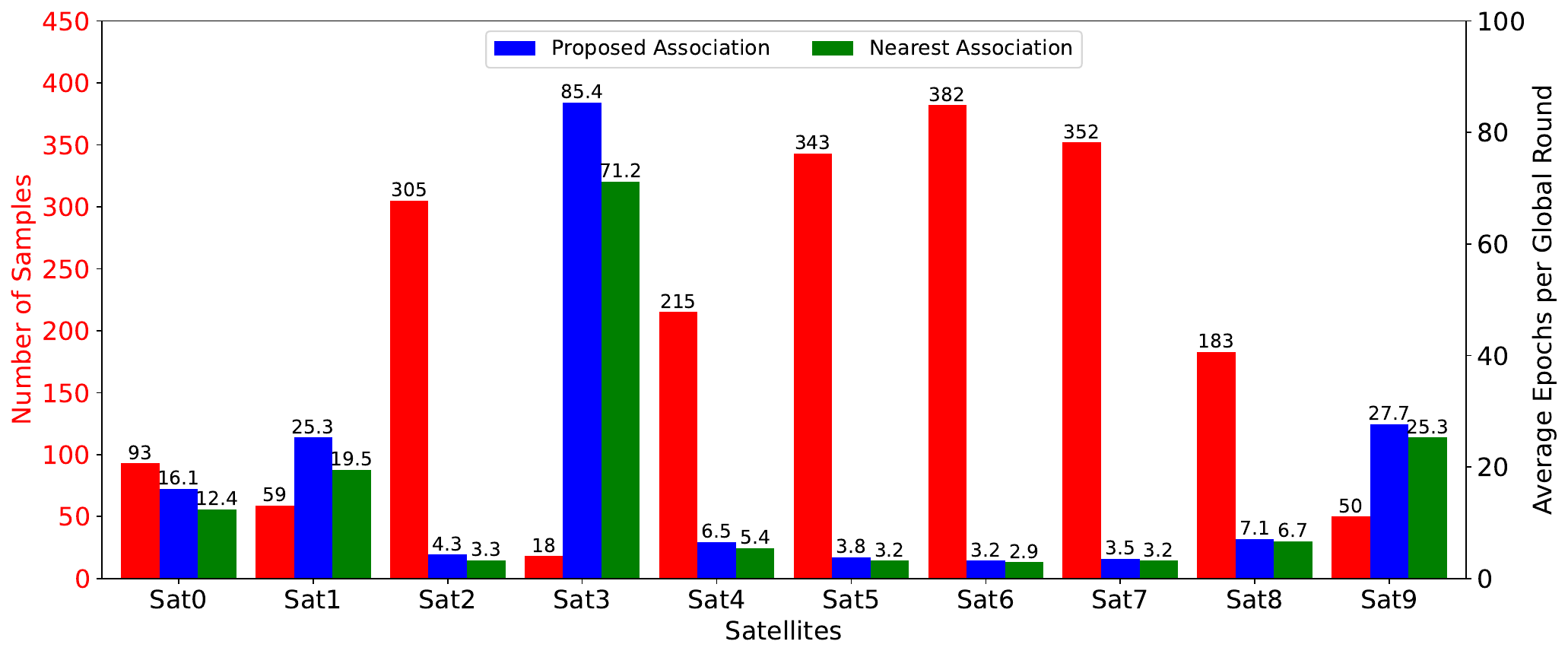}
    \caption{A comparison of the average training epochs per global round for each satellite across two distinct association approaches.}
    \label{AssociationResult}
\end{figure*}
\setcounter{figure}{3}
\begin{figure}[t]
    \centering
    \includegraphics[width=0.5\textwidth]{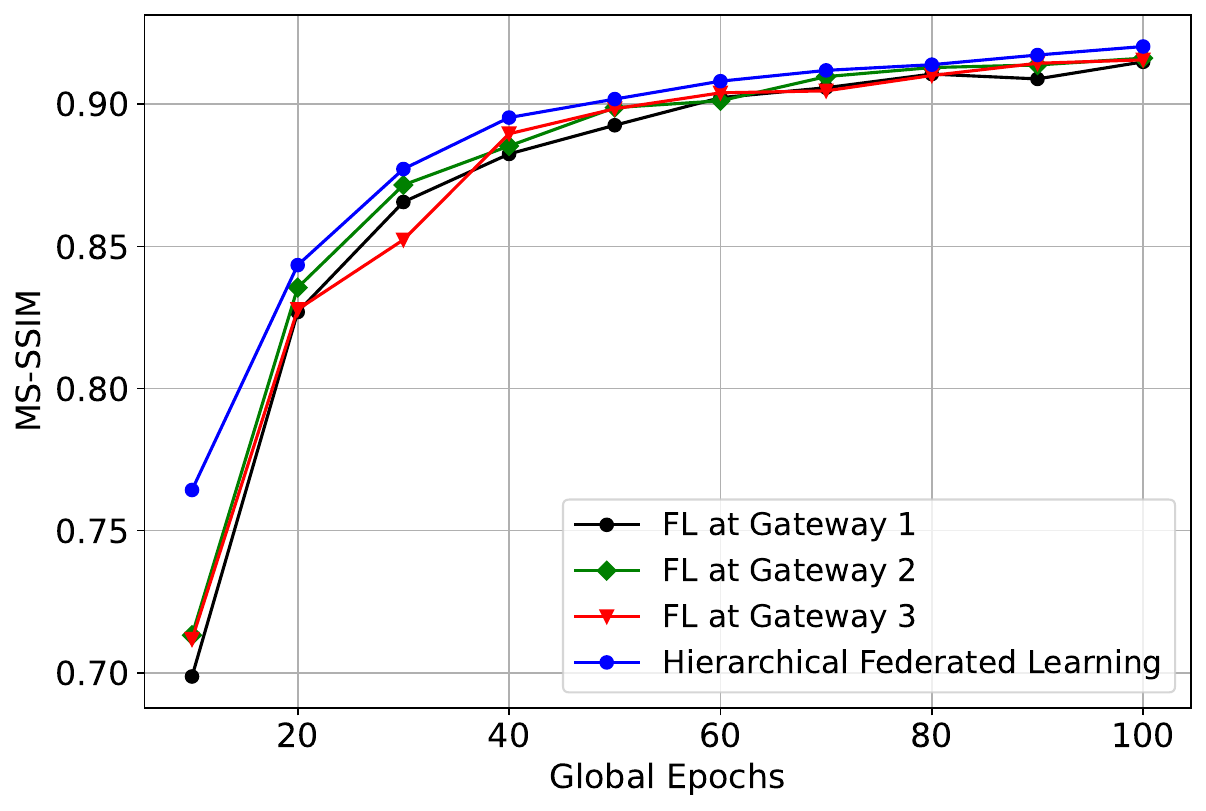}
    \caption{The convergence line of HFL against single gateway FL scenarios for the MS-SSIM metric.}
    \label{HFLPerformance-MSSIM}
\end{figure}

As observed in Fig.~\ref{HFLPerformance}, the learning models at the gateways exhibit rapid convergence in the initial training rounds, but after a certain point, the performance plateaus and struggles to improve in the later epochs. Specifically, the models at Gateways 1, 2, and 3 fluctuate around a peak signal-to-noise ratio (PSNR) value of 27.3 and fail to surpass this threshold. In contrast, the HFL framework continues to improve beyond this threshold after the 70th training epoch, demonstrating superior performance. The aggregation of the global model at the cloud server allows exposure to diverse data distributions, enhancing the model's generalization capability compared to single-gateway aggregation.

While the PSNR metric evaluates the difference between the original and reconstructed images at the pixel level, the mean structural similarity index (MS-SSIM) provides insights into the high-level structural similarity of the image. As shown in Fig.~\ref{HFLPerformance-MSSIM}, the HFL framework outperforms conventional FL in terms of structural similarity. Additionally, the HFL framework exhibits smoother and more stable convergence throughout the global training epochs, an important attribute for practical deployment in real-world satellite communication systems.

\subsubsection{Effect of the Satellite Association Variable on Model Training}
In Table~\ref{AssociateTable}, we evaluate the performance of the proposed satellite association technique compared to the nearest association approach under two aggregation schemes within the HFL framework using three gateways. These aggregation schemes are \textit{FedAvg} and \textit{FedAvep}. In the \textit{FedAvg} scheme, the contribution of each satellite is calculated based on the number of training samples associated with each gateway or cloud aggregation. In contrast, in the \textit{FedAvep} scheme, the contribution is determined by the product of the number of training samples and the epoch count, as outlined in Eq.~\ref{subregionaggregation} with $\beta = 1$.

Overall, we observe a significant improvement in both the PSNR and MS-SSIM metrics when switching from the nearest association approach to our proposed solution. Specifically, for the \textit{FedAvg} scheme, the PSNR improves by $0.26$ at an SNR of $1$ dB and increases to $0.47$ at $11$ dB under more favorable wireless conditions. Similarly, the MS-SSIM metric shows an increase with our proposed association approach. A similar trend is observed in the \textit{FedAvep} scheme, where the performance gap increases from $0.19$ at $1$ dB to $0.42$ and $0.51$ at $5$ dB and $11$ dB, respectively. This improvement is attributed to better utilization of satellite resources, as evidenced by the higher number of training epochs achieved under the same energy constraints.

\begin{table}[t]
    \centering
    \renewcommand{\arraystretch}{1.15} 
    \caption{Comparison of Average Computing Frequency and Connection Time Between Proposed and Nearest Association}
    \begin{tabular}{@{}ccccc@{}}
        \toprule
        \textbf{Satellite} & \textbf{Proposed (GHz, s)} & \textbf{Nearest (GHz, s)} \\ \midrule
        Sat 0 & (0.3664, 438) & (0.4197, 300) \\
        Sat 1 & (0.3708, 426) & (0.4256, 294) \\
        Sat 2 & (0.3732, 410) & (0.4236, 291) \\
        Sat 3 & (0.3636, 447) & (0.4030, 343) \\
        Sat 4 & (0.3741, 427) & (0.4133, 335) \\
        Sat 5 & (0.3768, 413) & (0.4109, 333) \\
        Sat 6 & (0.3931, 396) & (0.4127, 357) \\
        Sat 7 & (0.4054, 391) & (0.4221, 356) \\
        Sat 8 & (0.4024, 389) & (0.4174, 360) \\
        Sat 9 & (0.4000, 386) & (0.4207, 342) \\ \midrule
        \textbf{Avg} & (0.3856, 412) & (0.4169, 331) \\
        \bottomrule
    \end{tabular}
    \label{frequency_comparison}
\end{table}

\begin{table*}[t]
\caption{The performance of the proposed aggregation approach against other benchmarks under different levels of channel noise.}
\renewcommand{\arraystretch}{1.25} 
\begin{tabular}{|c|cc|cc|cc|cc|cc|}
\hline
Aggregation Mechanisms & \multicolumn{2}{c|}{FedAvg}         & \multicolumn{2}{c|}{FedAvep}           & \multicolumn{2}{c|}{FedIndi}           & \multicolumn{2}{c|}{FedLol}            & \multicolumn{2}{c|}{FedSEL}            \\ \hline
Metrics                & \multicolumn{1}{c|}{PSNR} & MS-SSIM & \multicolumn{1}{c|}{PSNR}    & MS-SSIM & \multicolumn{1}{c|}{PSNR}    & MS-SSIM & \multicolumn{1}{c|}{PSNR}    & MS-SSIM & \multicolumn{1}{c|}{PSNR}    & MS-SSIM \\ \hline
SNR=1 dB               & \multicolumn{1}{c|}{26.2396}  & 0.8691     & \multicolumn{1}{c|}{26.4792} & 0.8754  & \multicolumn{1}{c|}{26.2869} & 0.8699  & \multicolumn{1}{c|}{26.1658} & 0.8658  & \multicolumn{1}{c|}{\textbf{26.5537}} & \textbf{0.8773}  \\ \hline
SNR=3 dB               & \multicolumn{1}{c|}{27.0906}  & 0.8993     & \multicolumn{1}{c|}{27.4338} & 0.9056  & \multicolumn{1}{c|}{27.1892} & 0.9002  & \multicolumn{1}{c|}{27.0594} & 0.8973  & \multicolumn{1}{c|}{\textbf{27.4709}} & \textbf{0.9066}  \\ \hline
SNR=5 dB               & \multicolumn{1}{c|}{27.7784}  & 0.9203     & \multicolumn{1}{c|}{28.1848} & 0.9257  & \multicolumn{1}{c|}{27.9322} & 0.9223  & \multicolumn{1}{c|}{27.7492} & 0.9184  & \multicolumn{1}{c|}{\textbf{28.2221}} & \textbf{0.9267}  \\ \hline
SNR=7 dB               & \multicolumn{1}{c|}{28.2838}  & 0.9342     & \multicolumn{1}{c|}{28.7330}  & 0.9388  & \multicolumn{1}{c|}{28.4996} & 0.9375  & \multicolumn{1}{c|}{28.2569} & 0.9323  & \multicolumn{1}{c|}{\textbf{28.7840}}  & \textbf{0.9400}    \\ \hline
SNR=9 dB               & \multicolumn{1}{c|}{28.6300}  & 0.9429     & \multicolumn{1}{c|}{29.1072} & 0.9470   & \multicolumn{1}{c|}{28.9016} & 0.9471  & \multicolumn{1}{c|}{28.6119} & 0.9415  & \multicolumn{1}{c|}{\textbf{29.1735}} & \textbf{0.9481}  \\ \hline
SNR=11 dB              & \multicolumn{1}{c|}{28.8552}  & 0.9480     & \multicolumn{1}{c|}{29.3452} & 0.9520   & \multicolumn{1}{c|}{29.1654} & \textbf{0.9529}  & \multicolumn{1}{c|}{28.8483} & 0.9473  & \multicolumn{1}{c|}{\textbf{29.4237}} & 0.9527  \\ \hline
\end{tabular}
\label{AggregationTable}
\end{table*}

As depicted in Fig.~\ref{AssociationResult}, we present the number of training samples available at each satellite and the corresponding number of training epochs for both the proposed and nearest association approaches. A general observation reveals an inverse relationship between the number of training samples and the number of training epochs. Specifically, satellites with a higher number of training samples require more time and energy to complete a single training epoch. Conversely, satellites with fewer samples can train for more epochs, thereby better utilizing the available resources. With our proposed association approach, we achieve a higher number of training epochs across all satellites compared to the nearest association approach, resulting in improved performance, as illustrated in Table~\ref{AssociateTable}.

Furthermore, we compare the computing frequency between the two schemes, where it is evident that satellites under the nearest association approach require a higher computing frequency for network training. This is due to the shorter connection time between the satellites and the gateways, as shown in Table~\ref{frequency_comparison}. As indicated in equation (\ref{energycons}), energy consumption is quadratic to the computing frequency, which causes a faster depletion of the satellite's battery in the nearest association scheme. In contrast, our proposed association method ensures a longer connection time between the satellite and gateway, leading to a lower computing frequency, as per equation (\ref{ComputingFrequency}). This results in more efficient energy usage and enables a greater number of training epochs.

\subsubsection{Performance Achieved via Proposed Aggregation Method}  
In contrast to conventional methods, our framework effectively utilizes the computational resources of individual satellites by dynamically adjusting the number of training epochs based on their orbital trajectories and available energy. This adaptability makes the proposed approach unique. In this section, we evaluate various aggregation methods. As shown in Table~\ref{AggregationTable}, we compare several techniques for aggregating the global model at the gateways, including \textit{FedAvg}, \textit{FedAvep}, and \textit{FedLol}, the latter being a specialized technique for sub-region aggregation at the gateway. Additionally, \textit{FedIndi} treats training loss, training samples, and training epochs independently.

Overall, our proposed aggregation method outperforms the others across both metrics. Specifically, the method based solely on the loss function, \textit{FedLol}, performs the worst. Under an SNR of 1 dB, our method improves the average PSNR by up to $0.3879$ compared to \textit{FedLol} and $0.314$ compared to \textit{FedAvg}, with these performance gaps widening as wireless channel conditions improve. The inferior performance of \textit{FedLol} contrasts with findings in \cite{10531097}, which can be attributed to differences in the number of training rounds across clients. Satellites with fewer training samples undergo more training rounds, which helps reduce their loss but also increases the risk of overfitting to their limited data. As a result, these clients exert more influence during model aggregation, potentially degrading overall training performance.

While the differences in the MS-SSIM metric may initially appear minor, they become significant when considering the metric's maximum value of 1. Even under challenging wireless channel conditions at SNR = 1 dB, all aggregation methods exhibit high metric values and demonstrate improved performance as the noise level decreases.

\begin{table}[t]
    \centering
    \renewcommand{\arraystretch}{1.15} 
    \caption{Number of Training Samples per Satellite and Average Training Epochs under Different Dirichlet Distribution Scenarios}
    \resizebox{\columnwidth}{!}{
    \begin{tabular}{cccc}
        \toprule
        \textbf{Dirichlet Distribution} & \textbf{$\lambda = 0.1$} & \textbf{$\lambda = 1$} & \textbf{$\lambda = 10$} \\
        \toprule
        \textbf{Satellite} & \#Data, \#Epochs & \#Data, \#Epochs & \#Data, \#Epochs \\
        \midrule
        Sat 0  & (93, 16.1)  & (283, 4.9) & (191, 7.6) \\
        Sat 1  & (59, 25.3)  & (191, 7.4) & (193, 7.4) \\
        Sat 2  & (305, 4.3) & (128, 11.0) & (202, 6.9) \\
        Sat 3  & (18, 85.4)  & (195, 7.4) & (209, 6.9) \\
        Sat 4  & (215, 6.5) & (136, 10.6) & (181, 7.8) \\
        Sat 5  & (343, 3.8) & (156, 8.9) & (195, 7.0) \\
        Sat 6  & (382, 3.2) & (234, 5.5) & (195, 6.7) \\
        Sat 7  & (352, 3.5) & (221, 5.8) & (202, 6.4) \\
        Sat 8  & (183, 7.1) & (294, 4.3) & (185, 7.1) \\
        Sat 9 & (50, 27.7) & (162, 8.2) & (247, 5.2) \\
        \bottomrule
    \end{tabular}
    }
    \label{TableDistribution}
\end{table}

\setcounter{figure}{5}
\begin{figure}[t]
    \centering
    \includegraphics[width=0.49\textwidth]{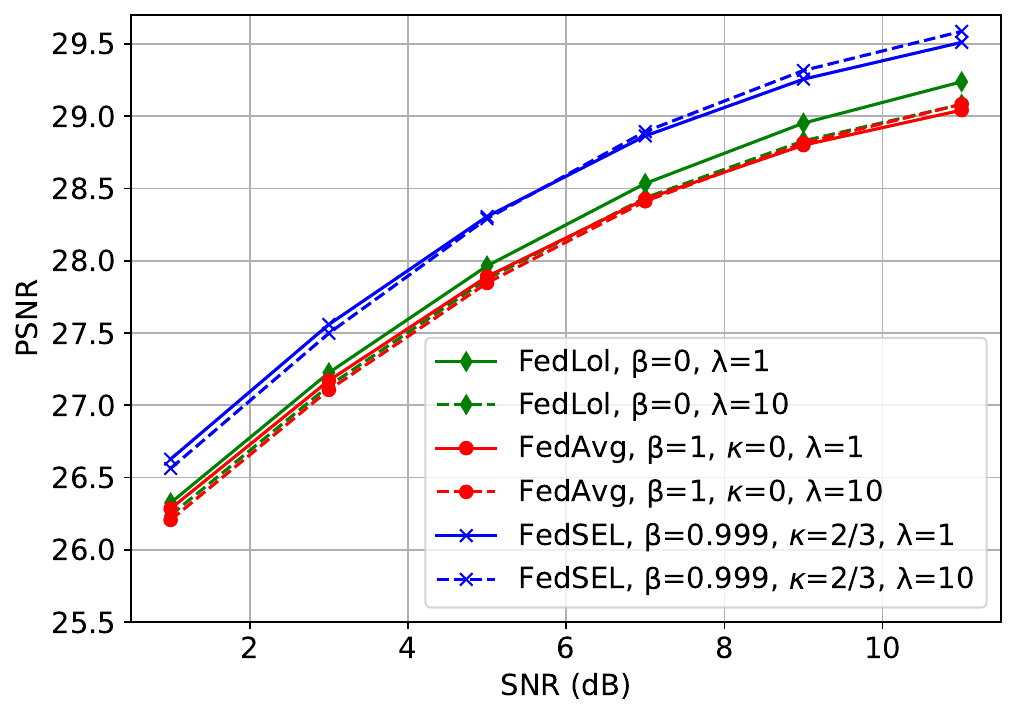}
    \caption{PSNR Performance Comparison of Aggregation Methods Under Varying Satellite Data Distributions}
    \label{distribution_PSNR_Figure}
\end{figure}

\subsubsection{Performance of the Proposed Framework under Different Data Distributions}  
Up to this point, we have conducted simulations using $\lambda = 0.1$, resulting in a highly imbalanced data distribution across satellites. To further assess the effectiveness of our proposed aggregation method, we also performed experiments with higher values of $\lambda$, specifically 1 and 10, where the data is distributed more evenly. While most existing works on FL in satellite networks focus on image classification tasks, our work addresses a reconstruction task. As a result, we do not consider class imbalance across satellites but rather focus on the number of samples available for each satellite. Table~\ref{TableDistribution} provides the details of the number of training samples and epochs for each satellite as $\lambda$ increases from $0.1$ to $1$, and up to $10$. Notably, when $\lambda = 10$, each satellite client has approximately $200$ samples.

As shown in Fig.~\ref{distribution_PSNR_Figure}, we compare the PSNR performance of the proposed aggregation method with \textit{FedAvg} and \textit{FedLol} under different data distributions across satellites. Our proposed method consistently outperforms the other benchmarks by large margins for all channel conditions and both values of $\lambda$, with \textit{FedLol} showing the second-highest performance. In general, all aggregation methods exhibit stable performance across varying data distributions in the satellites. This stability can be attributed to the nature of the image reconstruction task, where the objective is to minimize the MSE between the reconstructed and original images, rather than to achieve a definitive classification. Hence, performance improvements are measured in terms of MSE minimization rather than achieving a perfect match, as seen in classification tasks.

\begin{figure}[t]
    \centering
    \includegraphics[width=0.49\textwidth]{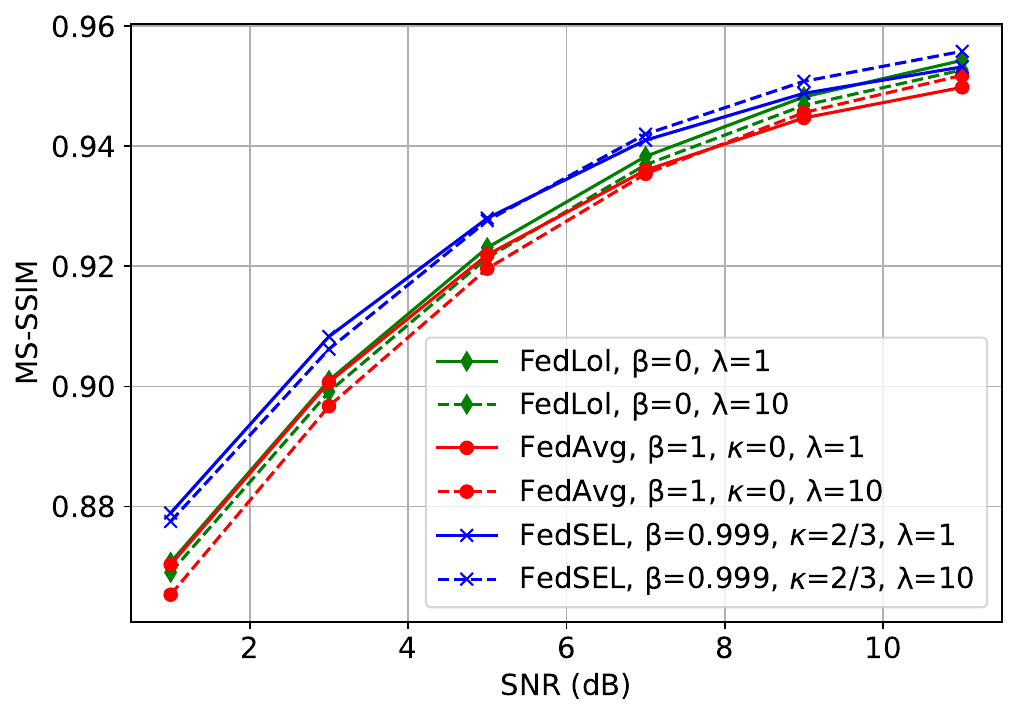}
    \caption{MS-SSIM Performance Comparison of Aggregation Methods Under Varying Satellite Data Distributions}
    \label{Distribution_MS-SSIM_Figure}
\end{figure}

Lastly, we present the performance differences in terms of the MS-SSIM metric for various channel conditions in Fig.~\ref{Distribution_MS-SSIM_Figure}. As observed, the performance gap between FedESL and FedAvg is significant at low SNR values, with a difference of approximately $0.009$. However, this gap decreases as the wireless channel conditions improve, narrowing to around $0.004$ in MS-SSIM value. Overall, all aggregation techniques demonstrate improved performance for the reconstructed images as the noise level decreases, effectively reflecting the adaptability of the semantic communication system to varying channel conditions.

\section{Conclusion}\label{Conclusion}
In this paper, we investigate a hierarchical federated learning framework for training semantic communication models in satellite networks. LEO satellites are capable of collecting vast amounts of data through their onboard sensors, but privacy concerns prevent direct data sharing. In the proposed framework, satellites function as learning agents in the FL process, distributed gateways act as sub-region aggregators, and a cloud server performs global model aggregation. We formulate a joint optimization problem for satellite association and contribution mechanisms within this hierarchical structure, taking into account practical constraints such as limited communication windows, energy availability, and the computational resources of the satellites. To address these challenges, we propose a novel satellite association strategy that integrates both orbital characteristics and locations to maximize the communication window time from the associated satellite to the gateway, which enables more local training and enhances the training efficiency. Additionally, we introduce an aggregation mechanism that accounts for various factors, including the number of training samples, training epochs, and lastly, local loss value when aggregating sub-region models, thereby enhancing the training process of the deep learning model used in semantic communication. Finally, we conduct extensive simulations to demonstrate the effectiveness of our approach compared to existing benchmarks. The results show that our framework achieves higher reconstruction quality, improved convergence stability, and better utilization of satellite resources, validating its potential for real-world deployment.

\ifCLASSOPTIONcaptionsoff

\fi

\bibliographystyle{IEEEtran}
\bibliography{mybib}




\end{document}